
\magnification=1200
\def\rund#1{\left( #1 \right)}
\font\twelverm=cmr12 at 20pt
\font\eightrm=cmr12 at 10pt
\font\tenmat=cmbsy10 at 10pt
\font\tenms=msam10 at 10pt
\def\pro{\tenmat\char47}
\def\lp{\tenms\char46}
\def\gp{\tenms\char38}
\def\rmi{\twelverm}
\def\rmj{\eightrm}
\input psfig
\vskip 3 pc
\centerline{\rmi Polytropic gas spheres: An approximate analytic }
\centerline{\rmi solution of the Lane-Emden equation}
\vskip 1 pc
\centerline{F.K. Liu}
\vskip 6 pt

\centerline {International School for Advanced Studies, Via Beirut 2--4, 34014
Trieste, Italy}
\centerline{e-mail: fkliu@neumann.sissa.it}
\vskip 3 pc

\centerline{\bf ABSTRACT}
Polytropic models play a very important role in galactic dynamics and
in the theory of stellar structure and evolution. However, in general, the
solution of the Lane-Emden equation can not be given analytically but
only numerically. In this paper we give a good analytic approximate
solution of the Lane-Emden equation. This approximation is very good for
any finite polytropic index $n$ and for the isothermal case at a level
$<1\%$. We also give analytic expressions of the mass, pressure,
temperature, and potential energy as a function of radius.

\vskip 1 pc

{\bf Key words:} star:evolution, star:neutron, star:white dwarfs,
Galaxy:kinematics and dynamics, galaxy: structure

\vskip 5 pc
\centerline{Accepted for publication in M.N.R.A.S.}
\vskip 10pc
\centerline{Ref. \quad SISSA \quad $159/95/$A}
\vfill

\par\break
\vskip 12 pt
\hskip -20pt {\bf 1. Introduction}
\vskip 6 pt

As stellar encounters are little important for galaxies, clusters
of galaxies, or globular clusters (Binney \& Tremaine 1987), the
fundamental dynamics describing these systems is that of a
collisionless system. In the general case the collisionless Boltzmann
equation cannot be solved because it involves too many independent
variables. However, we can get certain exact solutions of the
collisionless Boltzmann equation for a subset of possible
stellar-dynamical equilibria. The system with a polytropic
state equation, corresponding to isotropic velocity dispersion
tensors, is one of them. In stellar structure and evolution theory,
the polytropic model also plays an important role (Chandrasekhar 1939).

For a polytropic system, the relation of pressure $P$ and density
$\varrho$ is given by
$$
P = K \varrho^\gamma \equiv K \varrho^{1 + {1 \over n}} \quad ,
\eqno(1)
$$
where $K$ is the polytropic constant, $\gamma$ is the adiabatic index,
and $n$ the polytropic index. $K$ is fixed in the degenerate gas
sphere (e.g. in a white dwarf or in a neutron star) and free in a
non-degenerate system. In galactic dynamics $n$ is
larger than $1 \over 2$ (Binney \& Tremaine 1987) which means that no
polytropic stellar system can be homogeneous. $n$ ranges from $0$ to
$\infty$ in the case of the theory of stellar structure and evolution
(Kippenhahn \& Weigert 1989, Chandrasekhar 1939). With the polytropic
relation (1), hydrostatic equilibrium, and Poisson's equation for
gravitational potential field, we derive the Lane-Emden equation
(see, for instance, Binney \& Tremaine 1987, Kippenhahn \& Weigert
1989) for spherical symmetry:
$$
{d^2\omega \over d\xi^2} + {2 \over \xi} {d\omega \over d\xi} = -
\omega^n \quad . \eqno(2)
$$
The dimensionless variables $\omega$ and $\xi$ are defined by
$$\eqalign{
&\xi = A r \quad , \qquad A^2 = {4\pi G \over (n+1) K} \varrho_c^{n-1 \over n}
\quad , \cr
&\omega = \rund{\varrho \over \varrho_c}^{1/n}
\quad , \cr} \eqno(3)
$$
where $\varrho_c$ is the density at the center of the sphere and {\it
G} the Newtonian gravitational constant. $\omega$ corresponds to the
dimensionless gravitational potential. We have excluded the isothermal
case $n = \infty$, which we will discuss in detail in a later section.

The Lane-Emden equation must be solved with the original central conditions:
$$
\omega (0) = 1 \quad , \qquad \rund{d \omega \over d \xi}_{\xi = 0} =
0 \quad , \eqno(4)
$$
which will ensure the regularity of the solution at the center. The
solution gives $\omega$ as a function of $\xi$. From $\omega$ we can get
the density profile $\varrho$. Only for the three values of $n = $ 0,
1, and 5 can the solution be given in analytic form. Apart from these
three cases the Lane-Emden equation has to be solved numerically. As
the Lane-Emden equation has a regular singularity at $\xi = 0$, we can
expand $\omega (\xi)$ in a power series as
$$
\omega(\xi) = 1 - {1 \over 6} \xi^2 + {n \over 120} \xi^4 + \ldots
 \eqno(5)
$$

{}From (2) and (4) we have ${2 \over \xi} {d\omega \over d\xi} =- {2
\over 3} = 2 {d^2\omega \over d\xi^2}$ at $\xi = 0$. In this paper, we
make the approximation of taking the second derivative term as
${\delta \over 2} {2 \over \xi} {d\omega \over d\xi}$ to obtain a good
approximate analytic solution of the Lane-Emden equation. We give our
analysis for the isothermal Lane-Emden equation in section 2,
comparing our results with numerical solution. The general case is dealt
with in section 3 followed by discussion and conclusion in section 4.

\vskip 12 pt
\hskip -20pt{\bf 2. Lane-Emden equation for isothermal sphere}
\vskip 6 pt

The Lane-Emden equation for an isothermal sphere (see Kippenhahn \& Weigert
1989, Binney \& Tremaine 1987) can be written as
$$
{d^2 \omega \over d \xi^2} + {2 \over \xi}{ d\omega \over d\xi} =
e^{-\omega} \quad . \eqno(6)
$$
where $\omega$ is related to the mass density at dimensionless radius
$\xi$ by
$$
\varrho = \varrho_c e^{- \omega} \quad , \eqno(7)
$$
and $\varrho_c$ is the density at the origin. It is easy to find
(see also Binney \& Tremaine 1987) that
$$
\omega = - \ln \rund{2 \over \xi^2} \quad , \eqno(8)
$$
is one of the solutions of Eq. (6). This solution describes a model
known as the
singular isothermal sphere. Unfortunately, the singular isothermal
sphere has infinite density at $\xi = 0$. To obtain a solution that is
well behaved at the origin, equation (6) has to be integrated
with the central conditions
$$
\omega (0) = 0 \quad , \qquad \rund{d \omega \over d \xi}_{\xi = 0} = 0 \quad
. \eqno(9)
$$
Its solution can not be given by analytic expressions but only
numerical computation. The Lane-Emden equation has a regular
singularity at $\xi = 0$. In order to understand the behaviour of the
solutions there, a power series expansion similar to (5) can be derived
and has to be used. For large radius where the effect of the central
conditions is very weak the solution should asymptotically approach
the singular isothermal solution. The isothermal sphere consisting of
an ideal gas has an infinite radius as well as an infinite mass.

At small radius, a useful approximation to $\varrho (\xi) $ is the
modified Hubble law (Binney \& Tremaine 1987), which was introduced
empirically by King (1962),
$$
\rund{\varrho \over \varrho_c} = {1 \over \rund{1 + {\xi^2 \over 9}}^{3
\over 2}} \quad . \eqno(10)
$$
Comparing this relation with a numerical solution of (6) one can say that
for $\xi$ {\lp} 5 the relative error is less than 5 \%. Expression (10)
does not fit the isothermal profile well at $ \xi$ {\gp} 9 as it
approaches asymptotically to a logarithmic slope $- 3$ and not $- 2$,
as is required by the isothermal profile.

Since $\rund{d^2\omega \over d\xi^2}_{\xi=0}=1/3 = {1\over2}\rund{{2
\over\xi} {d\omega\over d\xi}}_{\xi=0}$ at $\xi =0$ from (6),
it is reasonable to approximate the second derivative term in (6) with
$$
{1\over2}\rund{{2 \over\xi} {d\omega\over d\xi}}
$$
as our first approximation and to get a first differential equation
$$
{3 \over 2} {2 \over \xi}{ d\omega \over d\xi} = e^{-\omega} \quad
, \eqno(11)
$$
which can be integrated immediately, giving
$$
\omega_1 = \ln \rund{1 + {\xi^2 \over 6}} \quad , \eqno(12)
$$
and
$$
{\varrho_1 \over \varrho_c} = e^{- \omega_1} = {1 \over 1 + {\xi^2 \over
6}} \quad , \eqno(13)
$$
where we have used the condition (9) to determine the integration
constant. The subscript 1 indicates a first approximation.

When $\xi$ is very large, Eq. (12) becomes
$$
\omega_1 \sim \ln \rund{\xi^2 \over 6} \quad , \eqno(14)
$$
which has a very similar profile to (8) and the same
tangent as the numerical solution. Figure 1 shows the numerical
solution of equation (6) and our first approximation (13), which
shifted by $1/3$ approximately reaches the singular solution for
large $\xi$.

\vskip 1pc
\psfig{figure=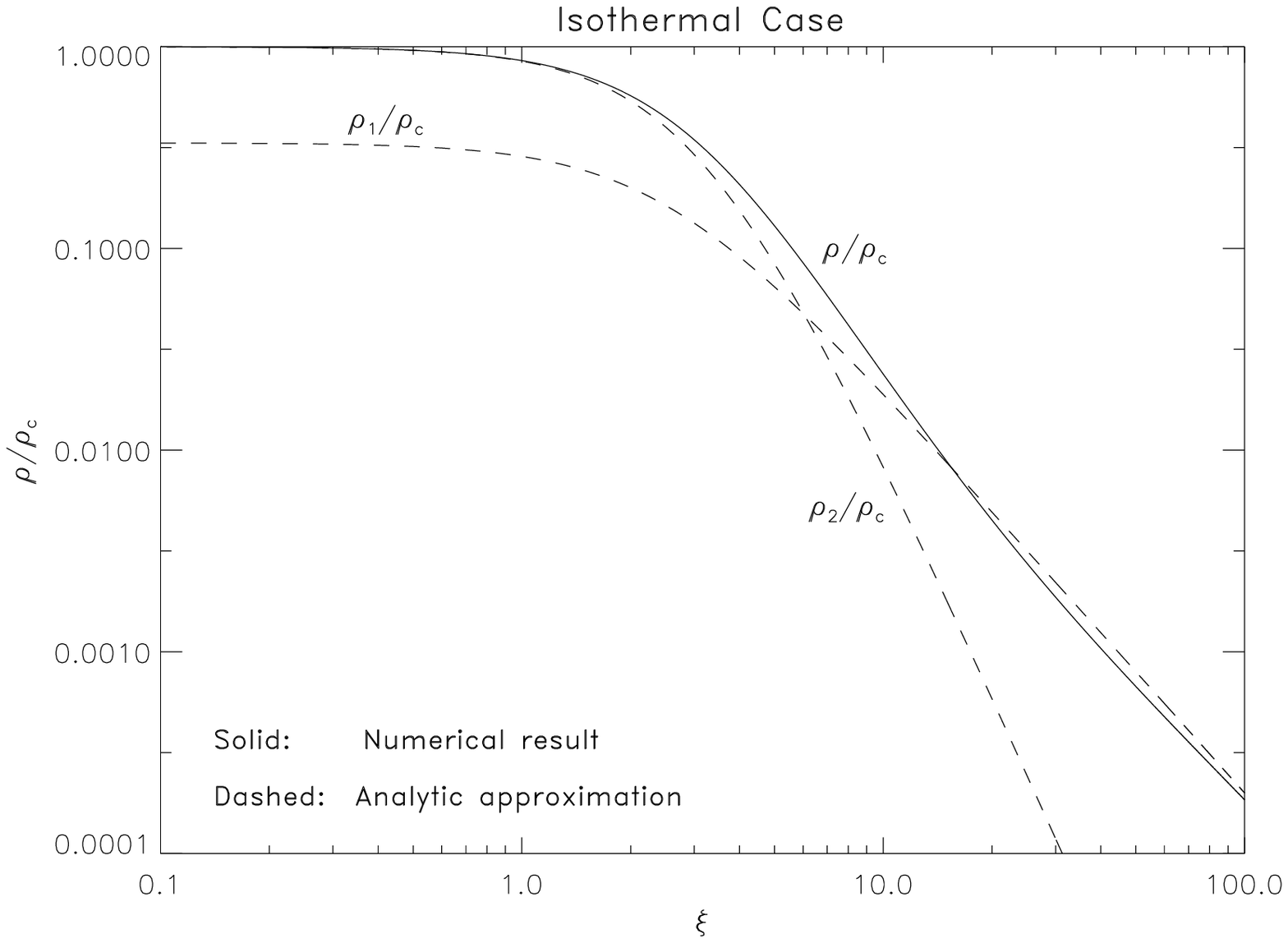,width=17cm}
{\rmj \hskip -20pt {\bf Fig. 1} The first order (dashed line indicated by
$\varrho_1/\varrho_c$) and the second order approximation (dashed line
indicated by $\varrho_2/\varrho_c$ ) of the density for the isothermal
case. The first order approximation has been multiplied by $1/3$ to
show its same tangent as the numerical result (solid line) for large
$\xi$.}

\vskip 2pc
Taking the second derivative of $\omega_1$ as our approximation to the
second derivative term and replacing the right hand side term in
equation (6) with (13), we obtain the second approximation to the solution
of equation (6)
$$
\omega_2 = 2 \ln\rund{1 + {\xi^2 \over 6}} - \rund{\xi^2 \over 6}
{1 \over 1 + {\xi^2 \over 6}} + cons \quad , \eqno(15)
$$
or
$$
{\varrho_2 \over \varrho_c} = e^{- \omega_2} = cons \rund{1 + {\xi^2
\over 6}}^{- 2} e^{\rund{\xi^2/6}/\rund{1 + {\xi^2 \over 6}}}
\quad , \eqno(16)
$$
where subscript 2 indicates the second approximation. Integration
constant $cons$ in (15) is determined by the central condition (9) and
is zero. Equations (16) as an approximation to the solution of
(6) is good for small $\xi$. For large $\xi$, the approximation has
an asymptotic behavior, $\varrho/\varrho_c$ {\pro} $\xi^{-4}$, which
does not fit the isothermal profile as the same
reason for the modified Hubble law discussed above. We also show the
second approximation in figure 1.

As the first order approximation (12) is good for large $\xi$ and the
second order approximation (15) is good for small $\xi$, we combine
them and obtain a more general approximation
$$
\eqalign{
{\omega_0} &= \alpha {\omega_2} + (1 - \alpha) {\omega_1} \cr
& =\rund{1+\alpha}\ln\rund{1 +{\xi^2\over6}} - {\xi^2\over6}
{\alpha \over 1 + {\xi^2 \over 6}} + cons \quad . \cr}
\eqno(17)
$$
where $0 \leq \alpha \leq 1$, coming from the condition (9), which
requires $\omega = 0$ at $\xi = 0$. For very large $\xi$, equation
(17) asymptotically approaches
$$
\rund{1+\alpha}\ln\rund{{\xi^2\over6}} - \alpha + cons,
$$
but not equation (8). So, we change the constant term in (17) to
$$
-\alpha\ln\rund{1 + \xi^2/A}
$$
to make equation (17) approach (8) for very large $\xi$. From (8) and
(17) A takes $3^{1/\alpha}12e$. Defining the general
approximation as $\omega = \omega_0 + \Delta\omega$ and substituting
it in equation (6), we get the final analytic approximation
$$
\eqalign{
{\omega} =& \rund{1+\alpha}\ln\rund{1 +{\xi^2\over6}} - {\xi^2\over6}
{\alpha \over 1 + {\xi^2 \over 6}} \cr
&-\alpha \ln\rund{1 + {\xi^2 \over 3^{1/\alpha} 12 e}} + \ln\rund{1 +
\rund{2^{-\alpha} -1}{D\xi^2 \over 1 + D\xi^2}} \quad , \cr}
\eqno(18)
$$
and
$$
{\varrho \over \varrho_c} = {\rund{1 + {\xi^2 \over 3^{1/\alpha} 12
e}}^\alpha \over  \rund{1 + {\xi^2 \over 6}}^{1 + \alpha} \rund{1 +
\rund{2^{-\alpha} -1}{D\xi^2 \over 1 + D\xi^2}}} e^{\alpha
\rund{\xi^2/6}/\rund{1 + {\xi^2 \over 6}}}
\quad , \eqno(19)
$$
where {\it e} is the natural constant, and {\it D} a constant
depending on $\alpha$. Equation (19) reaches the singular solution for
$\xi \rightarrow \infty$. The discrepancy of (19) to the numerical
solution is therefore at its largest value when $\xi$ is intermediate.
Adjusting $\alpha$ and {\it D}, we can modify the fitness for
intermediate $\xi$. One good combination is $\alpha = 0.551$ and $D =
3.84*10^{-4}$, which makes the largest relative error of equation (19)
to the numerical solution of (6) be 0.72\%. In figure 2, we compare the
structure of densities obtained both numerically and analytically and
show the relative error $(\varrho_n -  \varrho)/\varrho_n$, where
$\varrho_n$ is the numerical computation solution of (6). For $\xi
\rightarrow \infty$, the relative error approaches zero. As in the
isothermal case $\omega$ has no physical meaning, we give only the
result for density $\varrho$. The largest error for $\omega$ is also
about 0.72\%.

\vskip 1pc
\psfig{figure=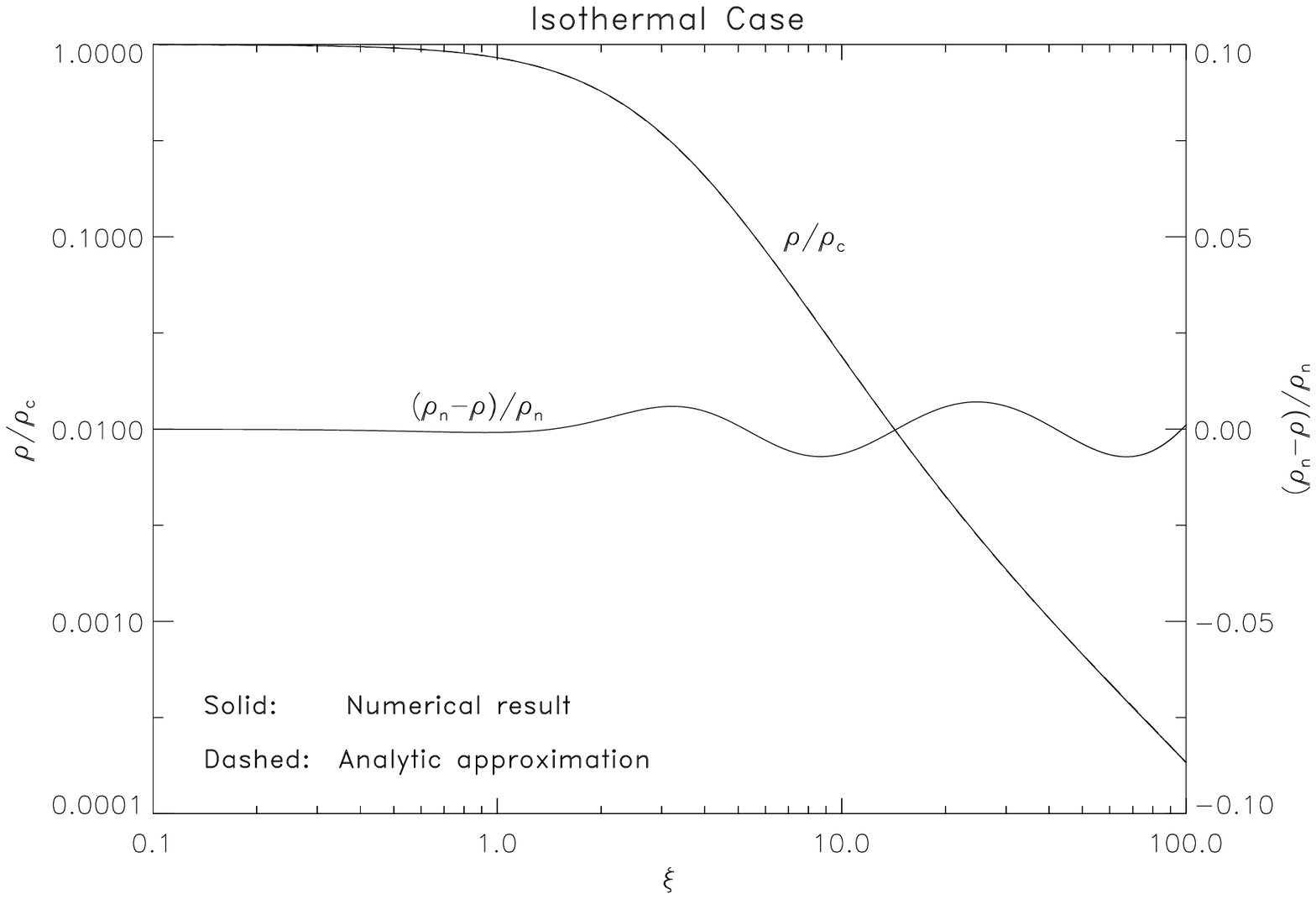,width=16cm}
{\rmj \hskip -20pt {\bf Fig. 2} The combination of the first and second
approximations with $\alpha = 0.551$ and $D= 3.84*10^{-4}$ (dashed
line) with its relative error to the numerical result (solid line)
for the isothermal case. It's difficult to distinguish the two lines
by eyes. The relative error is less than 0.72 \%.}

\vskip 2 pc
\hskip -20pt{\bf 3. Lane-Emden equation for general case }
\vskip 6 pt

We can apply a  similar method for the general case. We first discuss
the $n \not= 1$ case. The limit case $n = 1$ will be discussed later. For
$n \not= 1$, equation (2) also has a singular solution as in the
isothermal case
$$
\omega = \rund{ {\rund{1-n}^2 \over 2 \rund{n-3}} \xi^2}^{1 / (1-n)}
\quad . \eqno(20)
$$
When $n > 5$, the solution of equation (2) under central condition
(4) should asymptotically approach (20) for $\xi \rightarrow
\infty$. We approximately solve Eq. (2) with the same technique as
for the isothermal sphere. We approximate the second derivative term
in Eq. (2) with ${\delta \over 2} {2 \over \xi}
{d\omega \over d\xi}$ and get
$$
{2 + \delta \over 2} {2 \over \xi} {d\omega \over d\xi} = -\omega^n
\quad , \eqno(21)
$$
where $\delta$ is a constant. $\delta$ can affect the fitness of the
approximation strongly for $n<1$ but not for $n>1$. We can however
get very good approximation if fixing $\delta$ and adjusting other
parameters (see below) for any $n$. Therefore, we will always
take $\delta = 1$. Integrating equation (21), we get
$$
\omega_1 = \rund{1 + A_n \xi^2 }^{1/(1-n)}
\quad , \eqno(22)
$$
where $A_n$ is $\rund{n - 1}/6$. Here, we have used the central
condition (4) to determine the integration constant. However, this first
analytic approximation does not have the right behaviour for large
$\xi$ compared to the numerical solution of equation (2). Equations
(22) for any $n \geq 1$ can not reach zero at finite $\xi$ and give an
infinite radius. Integrating Eq. (2) gives
$$
\omega = - \int {\xi \over 2} \omega^n d\xi - \int {\xi \over 2}
{d^2\omega \over d\xi^2} d\xi  \quad . \eqno(23)
$$
Substituting (22) in the right-hand-side of Eq. (23) and integrating,
we get our second approximate analytic solution:
$$
\omega_2 =  cons + 2\rund{1 + A_n \xi^2}^{1 \over 1-n} + {\xi^2
\over 6} \rund{1 + A_n \xi^2}^{n \over 1 - n} \quad , \eqno(24)
$$
where the constant {\it cons} is determined by the central condition
(4) as $- 1$. This equation, however, reaches zero at finite $\xi$ for any
finite $n$ and gives a radius smaller than the one obtained
from numerical computation.

Since solution (22) gives too large values and (24) gives too small
values for large $\xi$, we construct a more general approximation for
$\omega$ as a linear combination of equations (22) and (24)
$$
\eqalign{
\omega_0& = \alpha \omega_2 + (1 - \alpha) \omega_1 \cr
&= -\alpha + \rund{1 + \alpha}\rund{1 + A_n \xi^2}^{1/\rund{1
-n}} + {\alpha \xi^2 \over 6}\rund{1 + A_n\xi^2}^{n
/\rund{1-n}} \quad . \cr} \eqno(25)
$$
We would like to point out that if we keep $\delta$ in (21) as a free
parameter, then equation (25) gives a good
approximation to the solution of (2) at level $<1\%$ with proper
$\delta$ and $\alpha$ for $n<1$. In this case $\delta$ varies from 0 to 1.
We notice that (25) asymptotically reaches the constant $-\alpha$ not
zero when $\xi \rightarrow \infty$ for $n \geq 5$. As (25) should
approaches the singular solution at very large $\xi$ (see also,
Eggleton, 1995)
as in the isothermal case, we change the constant $-\alpha$ in Eq.
(25) to $ - \alpha\rund{1 + B_n \xi^2}^{1/(1-n)}$ and let (25) for
$\xi \rightarrow \infty$ equal to the singular solution (20) when
$n >> 5$ and to zero when $n = 5$. We get
$$
B_n =\cases{ A_n\alpha^{n-1}\rund{1 + \alpha{n \over n-1} -
\rund{2\rund{n - 5} \over 9\rund{n + 1}}^{1/\rund{n -1}}}^{1 - n}
\quad , &for $n\ge 5$ \quad ; \cr
{n \rund{n-1} \over \rund{n + 1}^2} {6 \over 5 } \rund{4 \alpha
\over 4 + 5\alpha}^4 \quad , &for $n<5$ \quad; \cr} \eqno(26)
$$
where we have considered the conditions that $B_n$ should equal
zero at $n= 0$ and 1 and continue at $n = 5$ to get the expression
for $n < 5$. With $B_n$ given by (26), (25) gives a good
approximation with relative error less than $5\%$ for $n$ {\lp} 5 and
about $1\%$ for $n>>5$.
\vskip 3pt
\psfig{figure=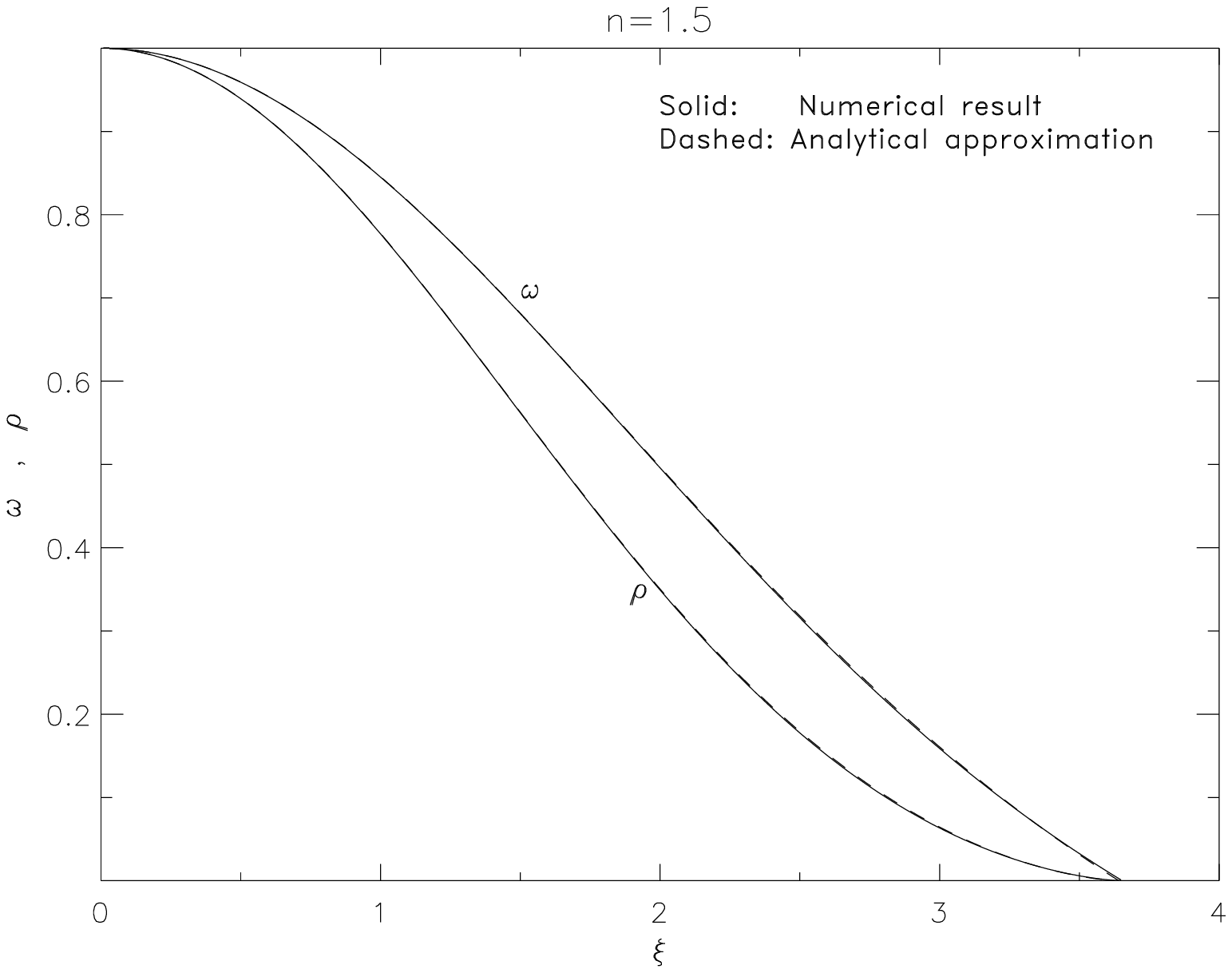,width=16cm}
{\rmj\hskip -20pt {\bf Fig. 3} The analytic approximation before a
$\Delta\omega$ modification given by equation (29) is fairly good.
Here is the result for $n=3/2$ as an
example. For $n=3/2$ case, the relative error is less than about 1\%
when $\alpha$ is taken as 0.44. }
\vskip 2pc

\hskip -20pt Figure 3 gives the results for $n=1.5$ as
an example. For $n=1.5$, the relative error is less than about $1\%$
when $\alpha = 0.44$.

So, if our general solution
is
$$
\omega = \omega_0 + \Delta\omega \quad , \eqno(27)
$$
$\Delta\omega$ would be very small. From Eqs.(2) and (25) as well as
(26), we get at very large $\xi$ for $n = 5$ and at $\xi
\rightarrow \xi_n$ for $n < 5$,
$$
\Delta\omega \sim {1 \over \xi} \quad , \eqno(28)
$$
as where $\omega_0 \sim 0$. In order to avoid the singularity
of $\Delta\omega$ at $\xi = 0$, we set
$$
\Delta \omega = {C_n \xi^{2\beta -1} \over
\rund{1 + D_n \xi^\beta}^2 }\quad , \eqno(29)
$$
where $C_n$ and $D_n$ are small constants and the index $\beta$ determined
from Eq.(2) is
$$
\eqalign{
\beta=&6.47- 7.01\beta_1 + 5.53\beta_1^2
 - 25.63 \beta_2 + 49.42\beta_2^2 - 26.88\beta_2^3 \quad , \cr }\eqno(30)
$$
where $\beta_1 = 1/\rund{1 + \rund{n -5}^2}$ and $\beta_2 = 1/\rund{1 +
\rund{n - 3}^2}$. With the same procedure and the same reason as for $n
\leq 5$, we have for $n > 5$
$$
\Delta\omega \simeq {E_n F_n \xi^{\eta+ {2 \over 1-n}}
\over 1 + F_n \xi^\eta} + {C_n \xi^{2\beta -1} \over \rund{1 + D_n
\xi^\beta}^2 } \quad ,  \eqno(31)
$$
where the index $\eta$ is $\rund{n-2}/\rund{n-4}$ and the coefficient
$E_n$ determined from equation (2) is
$$
E_n \simeq { 1 \over n -4} A_n^{1 \over 1 - n}
\rund{2 \rund{n - 5} \over 9 \rund{n + 1}}^{5 -n \over n -1} \rund{
\rund{n - 3 \over 3 \rund{n -1}}^{n -4 \over n -1} - \rund{2\rund{n
- 5} \over 9 \rund{n + 1}}^{n - 4 \over n - 1}} \quad,  \eqno(32)
$$
with which our approximation always asymptotically approaches the
singular solution (20) at large $\xi$ for $n>5$. To fit the solution
for intermediate $\xi$, the parameter $F_n$ can be given as
$$
F_n \simeq \cases{{\rund{n-5}^6 \over \rund{n-3}^5 \rund{4 n + 50}} ,
&for $n \ge 5$ \quad; \cr
0 , &for $n<5$\quad . \cr} \eqno(33)
$$

Finally, we configure a general result for any $n$
$$
\eqalign{
\omega =& -\alpha\rund{1 + B_n \xi^2}^{1/(1-n)} + \rund{1 + \alpha}
\rund{1+ A_n \xi^2}^{1/\rund{1 -n}} \cr
&+ {\alpha \over 6} \xi^2 \rund{1 + A_n\xi^2}^{n /\rund{1-n}}
+ {E_n F_n \xi^{\eta + {2 \over 1-n}} \over 1 + F_n \xi^\eta} +
{C_n \xi^{2\beta -1} \over \rund{1 + D_n \xi^\beta}^2 }
\quad , \cr} \eqno(34)
$$
where $\eta$ is $\rund{n-2}/\rund{n-4}$, $A_n$ is
$\rund{n-1}/6$, $B_n$ is given by (26), $E_n$ by (32) and $\beta$ by
(30). The free parameters $\alpha$, $F_n$, $C_n$ and $D_n$ should be
determined by fitting the numerical solution of equation (2). (34)
with proper values of the parameters $\alpha$, $C_n$, $D_n$ and $F_n$
can give a very good approximation solution of equation (2) under the
central condition (4).

For $n >> 5$, equation (34) tends to the singular
solution as $\xi \rightarrow \infty$. In this case, (34) always
approximates the solution of Eq.(2) with $ < 0.1 \%$. Especially when
$n>10$, the last term and when n {\gp} 500 the last two terms in
equation (34) are not important and can be ignored and the
approximation is within the level 0.1\%. As $n$ decreases, (34)
slowly deviates the singular solution to finally reach $\omega =
1 /(1 + \xi^2/3)^{1/2}$ at $\xi$ very large for $n=5$ where the
sum of the first three terms in (34) approach zero in accordance
with (25) and (26). (34) reaches zero at infinity for $n \geq 5$ and
at finite $\xi_n$ for $n < 5$. All the properties of (34) discussed
above are also the properties of the exact solution of (2) (a detail
dicussion to the general properties of the solution of equation (2),
see Binney \& Tremaine (1987), Kippenhahn \& Weigert (1989)).

\vskip 1pc
\psfig{figure=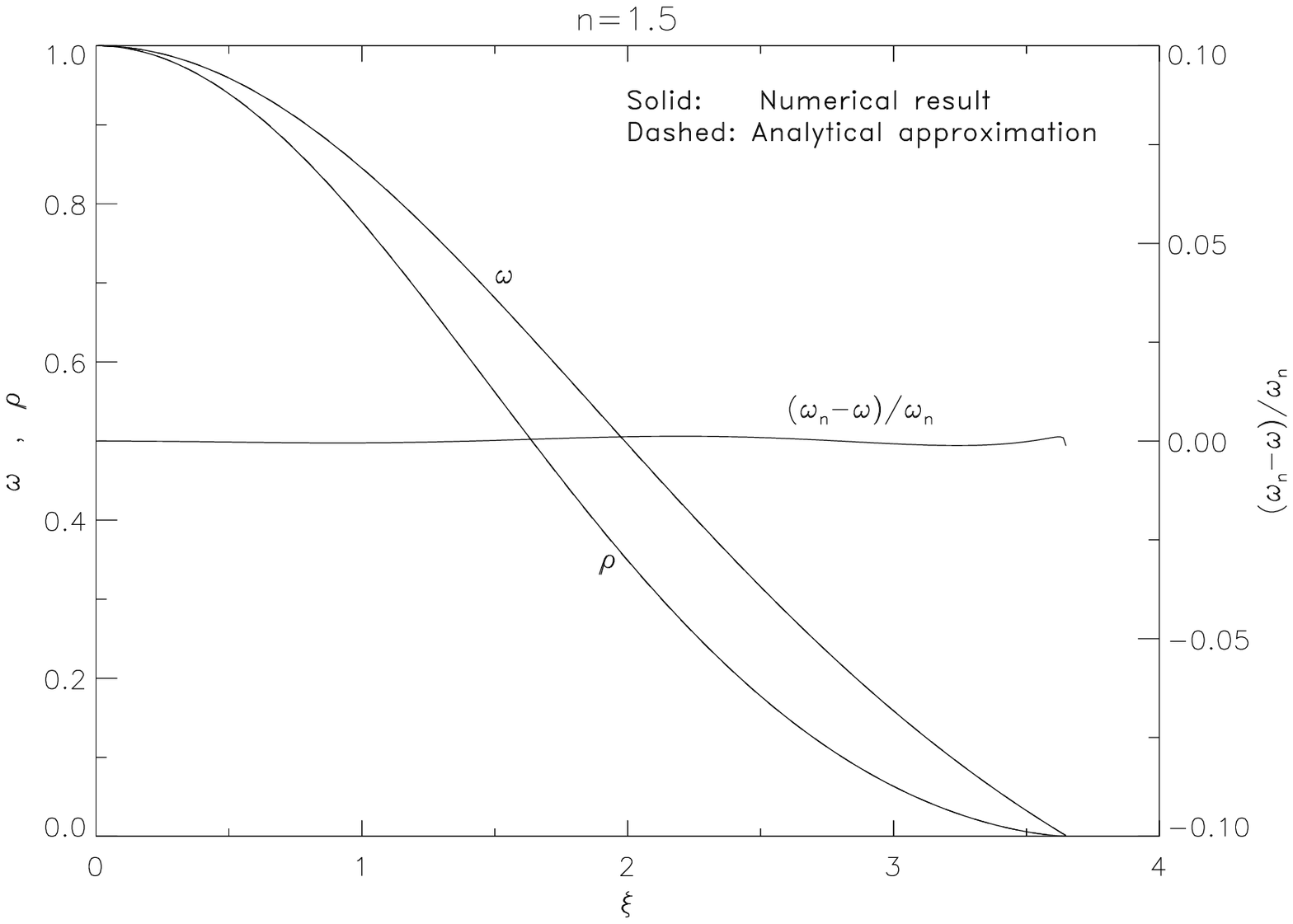,width=16cm}
{\rmj \hskip -20pt {\bf Fig. 4} Again for $n=3/2$ but after the modification
given by
equation (29). The figure shows the variation of potential $\omega$ and
density $\varrho$ with dimensionless radius $\xi$. The relative
approximation error of $\omega$ to numerical computation $\omega_n$ is also
shown. As in figure 2, it is difficult to distinguish the solid and
dashed lines. The largest error is 0.12\%}
\vskip 2pc

(34) is the exact solution for $n=0$. For general $n$, we have to adjust
$\alpha$, $C_n$, $D_n$ and $F_n$ to get a good approximation. Equation
(34) gives a good approximation to the solution of equation (2) for
small and large $\xi$, which is insensitive to the parameters within
proper ranges of values. For intermediate $\xi$, however, special
$\alpha$, $C_n$, $D_n$ and $F_n$ are needed to reduce the error of
the approximation. In figures 4 and 5, we show the variations of
potential $\omega$, density $\rho$ and relative error $(\omega_n -
\omega ) / \omega_n$ with radius $\xi$ for $n=3/2$ and for $n=5$,
respectively. For $n=3/2$, the relative error is larger near the
finite radius $\xi_n$ and oscillates around the mean value zero.
Such oscillation behavior also exists in figure 5 for $n=5$. In fact,
the oscillation always exists for any n.  We give the values of the
parameters used to get figures 4 and 5 and the largest
relative fitting errors $\epsilon_{max}$ in table 1. In table 1, we
give the values of the parameters for several polytropic indices
$n$ and the largest fitting errors $\epsilon_{max}$. We also give the
best fitting value for $F_n$ in the table. With the values of $\alpha$,
$C_n$ and $D_n$ and the value of $F_n$ given by equation (33), equation
(34) can also give an approximation at level $\epsilon_{max} < 1 \%$.

\vskip 1pc
\psfig{figure=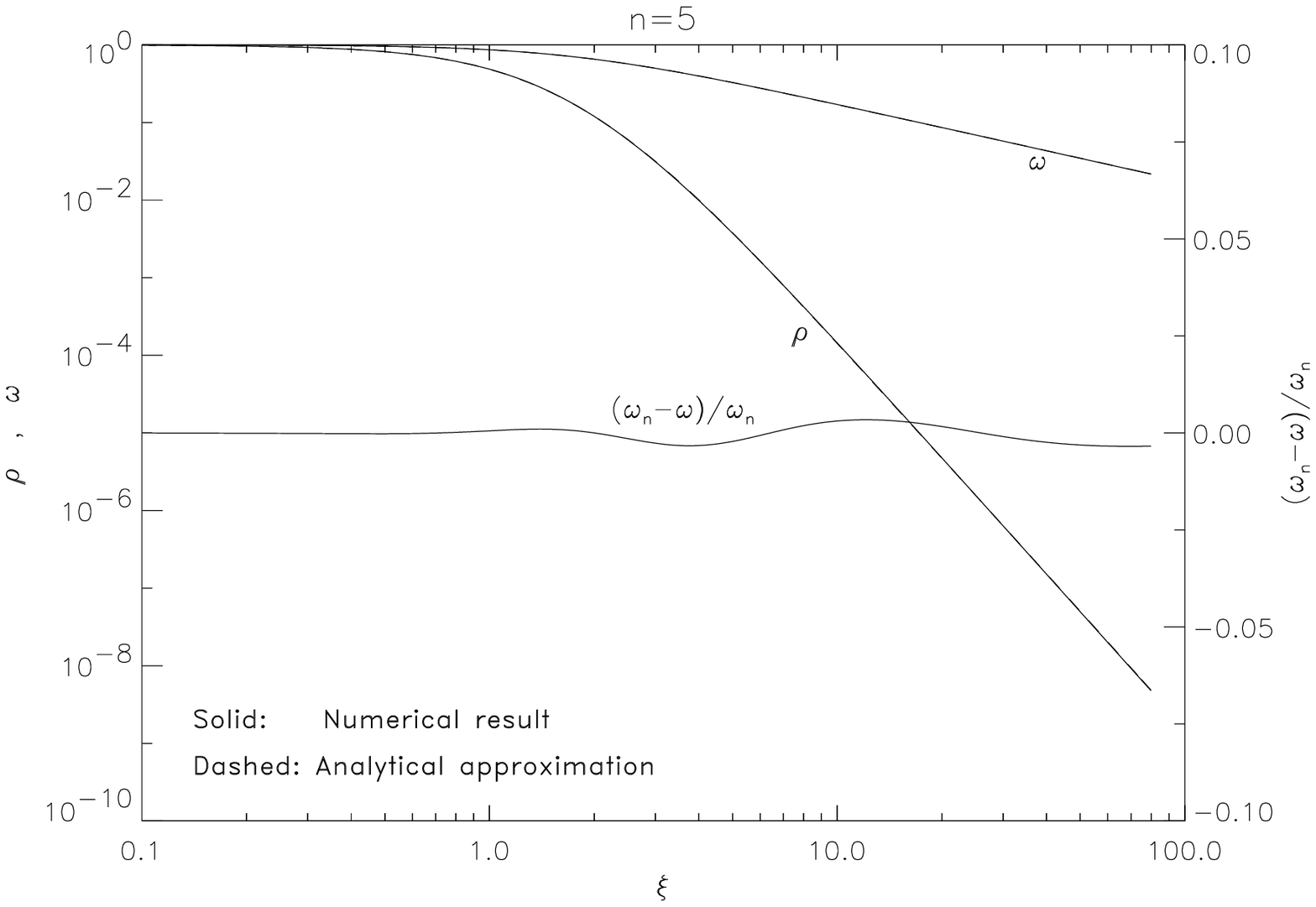,width=16cm}
{\rmj \hskip -20pt {\bf Fig. 5} Results for $n=5$. For $n \geq 5$, gas
sphere has an infinite radius and approaches its singular solution
when $\xi$ is very large. For $n=5$, the equation has an analytic
solution. The figure compares the approximate and the exact analytic
solutions and shows their relative difference. For $\xi \rightarrow 0$
or $\xi \rightarrow \infty$, the difference approaches zero. The same
as in the cases for $n=\infty$ and for $n=3/2$, it is very difficult to
distinguish the solid and dashed lines. The largest
difference is 0.34\%.}
\vskip 2pc

For $n < 5$, the relative error at $\xi \rightarrow \xi_n$ is larger
as the numerical solution $\omega_n$ approaches zero. (34) with
$B_n$ given by (26) reaches zero at finite $\xi$ and is the exact
solution for $n=0$. For $n \geq 5$, the error is at its maximum when
$\xi$ is in the range of around 1 to 20, depending on index $n$. When
$\xi$ is very large, the approximation approaches the singular
solution. With the increasing distance to the center of the gas
sphere, the effect of the central boundary (4) on the structure of
the sphere decreases. Therefore, the solution under the central
boundary condition (4) approaches the singular solution. The
approximation is very good even for $\xi \rightarrow \infty$. For
$n>10$, $C_n$ can be set to zero. For $n >500$, $E_n$ is less than
$10^{-3}$ and $\Delta\omega$ given by equation (31) becomes unimportant.

For $n = 1$, integrating (21) and considering the central boundary
condition (4) we get the first approximation
$$
\omega_1 = e^{- \xi^2 / 6} \quad . \eqno(35)
$$
Then, with the same procedure as for general n, we get our second
approximation

\par\break
{\rmj \hskip -20pt {\bf Table 1.} Parameters for several polytropic indices $n$

\hskip -20pt n polytropic index, $\alpha$ mixing parameter of the first and the
second approximations, $\epsilon_{max}$ (given in percent) maximum
relative error of analytic approximation to numerical calculation result.
For other parameters, see the text.
\vskip 3pt
\hrule height1pt
\vskip 1 pt
\hrule
\vskip  3 pt
\settabs 6 \columns
\+n &$\alpha$&$C_n$ &$D_n$& $F_n$& $\epsilon_{max}(\%)$\cr
\vskip 3 pt
\hrule
\vskip 3 pt
\+ 0.5&0.5&$1.30659*10^{-4}$&$2.3*10^{-4}$&0&0.37\cr
\+ 1&0.455&$1.27746*10^{-4}$&0&0&0.15\cr
\+1.5&0.481&$4.61841*10^{-4}$&$1.1*10^{-3}$&0&0.12\cr
\+ 2&0.512&$8.3218*10^{-4}$&$2.56*10^{-2}$&0&0.28\cr
\+ 3&0.53&$5.56215*10^{-4}$&$2.745*10^{-2}$&0&0.60\cr
\+ 5&0.545&$1.9415*10^{-3}$&$3.348*10^{-2}$&0&0.34\cr
\+ 6&0.56&$2.0*10^{-4}$&$3.0*10^{-2}$&$5.3*10^{-5}$&0.50\cr
\+ 7&0.526&$1.53*10^{-7}$&$1.1*10^{-3}$&$7.40*10^{-4}$&0.39\cr
\+ 8&0.540&$6.1*10^{-8}$&$1.0*10^{-3}$&$2.7*10^{-3}$&0.35\cr
\+ 9&0.546&$3*10^{-8}$&$9*10^{-4}$&$5.4*10^{-3}$&0.26\cr
\+10&0.538&$1.12*10^{-8}$&$9*10^{-4}$&$8.91*10^{-3}$&0.24\cr
\+ 20&0.613&$0$&$...$&$4.2*10^{-2}$&0.12\cr
\+ 50&0.65&$0$&$...$&$0.101$&0.060\cr
\+500&0.58&$0$&$...$&0.19&0.021\cr
\vskip 3 pt
\hrule
\vskip 1 pt
\hrule height1pt}
\vskip 2 pc

$$
\omega_2 = - 1 + 2 e^{-\xi^2 /6} + {\xi^2 \over 6} e^{- \xi^2 /6}
\quad . \eqno(36)
$$
(35) and (36) are also the limit of expressions (22) and (24) for $n
\rightarrow 1$, so the $\omega$ is continuous at $n = 1$. From the
limit case of equations (26) and (34) for $n = 1$, we therefore get
the final result for the $n = 1$
$$
\eqalign{
\omega =& -\alpha e^{-{3 \over 10} \rund{4\alpha \over 4 + 5\alpha}^4
\xi^2}+ \rund{1 + \alpha} e^{- \xi^2 \over 6} + {\alpha \over 6} \xi^2
e^{-\xi^2 \over 6} + {C_n \xi^{2\beta -1} \over \rund{1 + D_n \xi^\beta}^2 }
\quad ,
\cr} \eqno(37)
$$
where $\beta$ is given by equation (30), $C_n$ and $D_n$ are
constant. In table 1, we also give a set values of $C_n$ and $D_n$ and
the largest approximate error of equation (37). For $n=1$, the largest
relative error of the approximation is only $0.15\%$.

\vskip 1pc
\hskip -20pt{\bf 4. Discussion and conclusion}
\vskip 6 pt
We gave our analytic approximation solutions to the Lane-Emden equations
for isothermal sphere in section 2 and for general polytropic cases in
section 3. From the analytic results we can derive some useful
relations.

{}From (3) and (2) for the general case and from (7) and (6) for the
isothermal case, the mass contained in a sphere of radius $r$
$$
\eqalign{
m &= \int_0^r 4 \pi r^2 \varrho dr \cr
&= 4 \pi \varrho_c r^3 \rund{ \mp {1 \over \xi} {d\omega \over d\xi}}
\quad , \cr} \eqno(38)
$$
where $-$ sign is for the general case and $+$ for the isothermal case.
Dimensional $r$ and non-dimensional $\xi$ are related for the general
case by equation (3) and for the isothermal case by
$$
\xi = \rund{4\pi G \varrho_c \over K}^{1/2}r = A r \quad . \eqno(39)
$$
For the general case and $n \not= 1$, it follows from (34) that
$$
\eqalign{
m(r) =& 4 \pi \varrho_c  r^3 \cr
& \rund{{2\alpha B_n \over 1 -n}\rund{1 + B_n \xi^2}^{n \over 1 - n}
+ {1\over 3}\rund{1 + A_n\xi^2}^{n \over 1 - n} +
{n \alpha\over 18}\xi^2\rund{1 + A_n\xi^2}^{{n \over 1 - n} - 1 }} \cr
&- E_n F_n \xi^{\eta + {2 \over 1 - n} -2} {\eta
+{2\over 1 -n} + {2\over 1-n}F_n\xi^\eta \over \rund{1 +
F_n\xi^\eta}^2} - C_n\xi^{2\beta-3} {2\beta -1 -
D_n\xi^\beta \over \rund{1 + D_n\xi^\beta}^3} \quad . \cr } \eqno(40)
$$
For $n < 5$, the star has a finite radius $R$ or $\xi_n$, which is
estimated from (34) by setting $\omega$ to zero, and a finite mass
$M$. From (40) we can get the expression for mass $M$ from which
the central density $\varrho_c$ is determined. However,
for $n \geq 5$ the polytropic gas sphere has infinite radius and
infinite mass. In this case we cannot estimate the central density
directly. From equations (1)and (3) we have
$$
P = K \varrho_c^{n + 1 \over n} \omega^{n+1} \quad , \eqno(41)
$$
where $\omega$ is given by (34) and $K$ is determined by equation (3)
as $A$ is given by $A = \xi_n / R$ for $n < 5$. For an ideal gas the
temperature is given as
$$
T = {\mu m_p \over k} {P \over \varrho} = {\mu m_p \over k} K
\varrho_c^{1/n} \omega \quad , \eqno(42)
$$
where $\mu$ is the mean molecular weight, $m_p$ is the mass of atomic
Hydrogen, and $k$ is the Boltzmann constant. From the definition of
potential energy within radius $r$ of the sphere
$$
E_g \equiv - \int_0^m {G m' \over r'} dm' \quad , \eqno(43)
$$
the polytropic relation (1), and the hydrostatic equilibrium equation,
we get for $n \not= 5$
$$
E_g = - {3 \over 5 - n} {Gm^2 \over r} - {3 \over 5 - n} \Phi m - {n +
1 \over 5 - n} 4 \pi r^3 P \quad , \eqno(44)
$$
where $\Phi$ is gravitational potential, given by
$$
\Phi = - (n + 1) K \varrho^{1 \over n} \quad .  \eqno(45)
$$
This means that for $n < 5$ the polytropic sphere has finite radius
and the potential is set to zero at the surface. For $n \geq 5$ the
potential becomes zero at infinity.

In polytropic models, $n=3$ and $n=3/2$ are two important cases. We
have given $n=3/2$ as one example before. Here, as another
example, we construct a polytropic model of index 3 of the sun
($M=1.989 * 10^{33} g$ and $R = 6.96 * 10^{10} cm$) and compare
our results with that from numerical computation. For $n = 3$, from
table 1 and equations (26) and (30), we have $\alpha= 0.53$, $\beta=
2.199$, $B_n = 4.648*10^{-3}$, $C_n=5.56215*10^{-4}$ and $D_n =
2.745*10^{-2}$. Then we get $\xi_3 = 6.897$ from (34) and $A = \xi_3/R
= 9.895 * 10^{-11}$. Substituting these in equation (40) we get
central density $\varrho_c = 76.96$ g cm$^{-3}$, $K=3.87 * 10^{14}$
from equation
(3) and consequently, $P_c = 1.26 * 10^{17}$ dyn/cm$^2$ from equation
(41) as $\xi = 0$ at the center. For the ideal gas $\mu = 0.62$ we get
from equation (42) the central temperature $T_c = 1.2 * 10^7$ K. From
Kippenhahn and Weigert (1990), numerical computation gives $\xi_3 =
6.897$, $\varrho_c = 76.39$ g cm$^{-3}$, $P_c = 1.24 * 10^{17}$
dyn/cm$^2$ and $T_c = 1.2 *10^7$ K. We see that our approximation for
$n = 3$ gives a quite good result at a level 0.7\% for density, 1.6\%
for pressure and 0\% for temperature.

For the isothermal sphere, the mass from equations (18) and (38) is
$$
\eqalign{
m =& 8 \pi \varrho_c r^3 \cr
&\rund{{1 +\alpha \over 6 + \xi^2} -{6\alpha \over
\rund{6+\xi^2}^2} - {\alpha \over 3^{1/\alpha}12 e +\xi^2} + {D
\rund{2^{-\alpha} -1} \over \rund{1 + 2^{-\alpha} D \xi^2} \rund{ 1 +
D\xi^2}}} \quad , \cr} \eqno(46)
$$
where $\alpha = 0.551$ and $D=3.84*10^{-4}$. When $\xi >> \sqrt{6}$ the mass
changes with radius as $m$ {\pro} $r$. The pressure is given by
polytropic relation (1) for $n = \infty$. The constant $K$ is related
to the central density $\varrho_c$ by (39). However, as the isothermal
sphere does not have a finite
radius, the constant $A$ cannot be determined as in the case $n < 5$. In
the isothermal sphere the temperature is constant everywhere. From
definition (43) and the hydrostatic equilibrium equation, we find
the potential energy within radius $r$ for the isothermal sphere
$$
E_g = 4\pi r^3 K \varrho - 3 K m \quad , \eqno(47)
$$
where $\varrho$ is given by (19) and $m$ by (46).

In this paper we have given a good analytic approximate solution of
the Lane-Emden equation with which we have obtained analytic
expressions for the mass contained in radius $r$, the pressure,
temperature, and gravitational potential energy within radius $r$.
It would be interesting to apply the approximation in modeling the
structures of stars and galaxies.

\vskip 12pt
\hskip -20pt{\bf 5. Acknowledgments}
\vskip 6 pt

The author is grateful to Dr. A. Lanza for highlighting discussions
and for careful reading of the manuscript and indebted to the referee
Prof. P. Eggleton for his helpful comments and kindly sending of his
result. Many thanks are also given to Dr. R. Stark, Dr. D. Parashar
and Dr. M. Marengo for a careful reading of the manuscript.

\vskip12pt
\hskip -20pt {\bf References}
\vskip 6pt

\hskip -20pt Binney J., and Tremaine S. 1987, Galactic Dynamics,
Princeton: Princeton University Press

\hskip -20pt Chandrasekhar S. 1939, An Introduction to the theory of
Stellar Structure, Chicago: University of Chicago Press.

\hskip -20pt Eggleton, P. 1995, private communication

\hskip -20pt King I. R. 1962, AJ, 67, 471

\hskip -20pt Kippenhahn R., and Weigert A. 1990, Stellar Structure
and Evolution, Berlin: Springer-Verlag

\bye